\documentclass[final,authoryear,5p,times,twocolumn]{elsarticle}
\pdfoutput=1

\usepackage{graphicx}

\usepackage{amssymb}

\usepackage[pdftex,pdfpagemode={UseOutlines},bookmarks,bookmarksopen,colorlinks,linkcolor={blue},citecolor={green},urlcolor={red}]{hyperref}
\usepackage{hypernat}

\journal{Astronomy \& Computing}

\usepackage{upquote}

\usepackage{upgreek}

\usepackage{color}

\newcommand*\figref[1]{Fig.~\ref{#1}}

\begin{document}

\begin{frontmatter}



\title{Observatory/data centre partnerships and the VO-centric archive:
  The JCMT Science Archive experience}


\author[lsst]{Frossie Economou\corref{cor1}}
\ead{frossie@lsst.org}
\author[cadc]{S\'{e}verin Gaudet}
\author[cornell,jac]{Tim Jenness}
\author[jac]{Russell O.\ Redman}
\author[cadc]{Sharon Goliath}
\author[cadc]{Patrick Dowler}
\author[jac]{Malcolm~J.~Currie}
\author[jac]{Graham~S.~Bell}
\author[jac]{Sarah~F.~Graves}
\author[cadc]{John~Ouellette}
\author[jac,nrc,uvic]{Doug~Johnstone}
\author[cadc]{David Schade}
\author[uherts,jac]{Antonio~Chrysostomou}

\cortext[cor1]{Corresponding author}

\address[lsst]{LSST Project Office, 933 N.\ Cherry Ave, Tucson, AZ 85721, USA}
\address[cadc]{Canadian Astronomy Data Centre, National Research Council of Canada, 5071 West Saanich Road., Victoria, BC V9E 2E7, Canada}
\address[cornell]{Department of Astronomy, Cornell University, Ithaca,
  NY 14853, USA}
\address[jac]{Joint Astronomy Centre, 660 N.\ A`oh\=ok\=u Place, Hilo, HI
  96720, USA}
\address[nrc]{NRC-Herzberg Institute of Astrophysics, 5071 West Saanich Road,
Victoria, BC V9E~2E7, Canada}
\address[uvic]{ Department of Physics and Astronomy, University of Victoria, PO Box 3055 STN CSC, Victoria, BC V8W~3P6, Canada}
\address[uherts]{Centre for Astrophysics Research, University of Hertfordshire, College Lane, Hatfield, Hertfordshire AL10 9AB, UK}

\begin{abstract}

  We present, as a case study, a description of the
  partnership between an observatory (JCMT) and a data centre (CADC)
  that led to the development of the JCMT Science Archive (JSA). The
  JSA is a successful example of a service designed to use Virtual Observatory (VO)
  technologies from the start. We describe the motivation, process and
  lessons learned from this approach.

\end{abstract}


\end{frontmatter}


\newcommand{\mnras}{MNRAS}
\newcommand{\aap}{A\&A}
\newcommand{\aaps}{A\&AS}
\newcommand{\pasp}{PASP}
\newcommand{\apj}{ApJ}
\newcommand{\apjs}{ApJS}
\newcommand{\qjras}{QJRAS}
\newcommand{\an}{Astron.\ Nach.}
\newcommand{\ijimw}{Int.\ J.\ Infrared \& Millimeter Waves}
\newcommand{\procspie}{Proc.\ SPIE}
\newcommand{\aspconf}{ASP Conf. Ser.}


\newcommand{\ascl}[1]{\href{http://www.ascl.net/#1}{ascl:#1}}


\section{Origins}

The James Clerk Maxwell Telescope (JCMT) has collaborated with the Canadian Astronomy Data Centre (CADC) to
create the JCMT Science Archive (JSA) which provides raw and reduced JCMT data to the astronomical community
\citep{2008SPIE.7016E..16G,2008ASPC..394..450E,2008ASPC..394..135G,2011ASPC..442..203E}.
As a new generation of instruments was being developed for the JCMT in
the early 2000s \citep[HARP/ACSIS \& SCUBA-2;][]{2000ASPC..217...33D,2003SPIE.4855....1H},
it became clear that
the data rates from these instruments, of order 10\,MB/s, were going to be significantly
higher than earlier submillimeter instrumentation. In
particular SCUBA-2 was the first generation of submillimeter camera
that could be considered to be suitable for use as a large-scale
survey instrument. Exploratory discussions on the JSA between JCMT and CADC
began in 2003 and culminated in a decision to approve the
collaboration in May 2005 \citep{2005JCMTN23}. Development effort was
obtained in-house and also from the addition of two programmers recruited from
the UK Starlink project \citep{1982QJRAS..23..485D}, which had recently been closed.

The commitment to a JCMT Science Archive was followed shortly
afterwards by the approval of the JCMT Legacy Survey programme in July
2005 \citep{2005JCMTN23}. To ensure survey participation in the JSA
the JCMT Data Users' Group (JDUG) was created in early 2006 to provide
stakeholder input into the pipeline operation and advanced data
products \citep{2006JCMTN24R}.

\section{Motivation: Observatory}

\begin{figure}[t]
\includegraphics[angle=-90,width=\columnwidth]{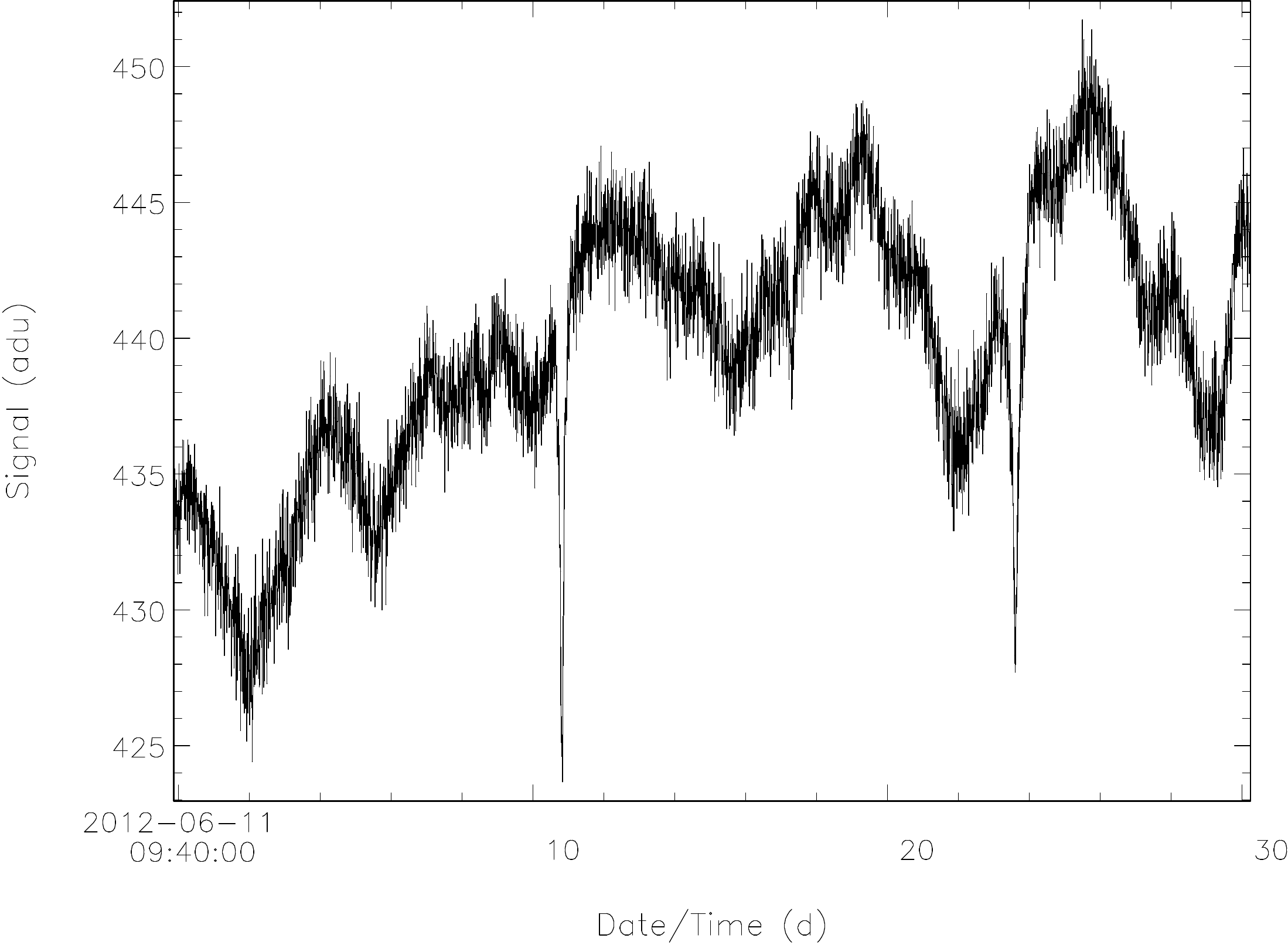}
\caption{Single bolometer time-series from a subset of a SCUBA-2
  observation of G34.3 from 2012 June 11th. The final image is shown in
  \figref{fig:g34}. The negative spikes are the detections of the
  bright central source.}
\label{fig:g34ts}
\end{figure}

Submillimeter data has traditionally been rather esoteric, closer to
radio than the optical/infrared regime familiar to most
astronomers. Raw data is typically in time series format (\figref{fig:g34ts}), and requires
in-house algorithms for transformation to science-ready formats such
as spectra or images. Calibration is
difficult due to the dominant and highly variable effect of the water
vapour in Earth's atmosphere
\citep[e.g.,][]{2002MNRAS.336....1A,2013MNRAS.430.2534D}.

\begin{figure}[t]
\includegraphics[width=\columnwidth]{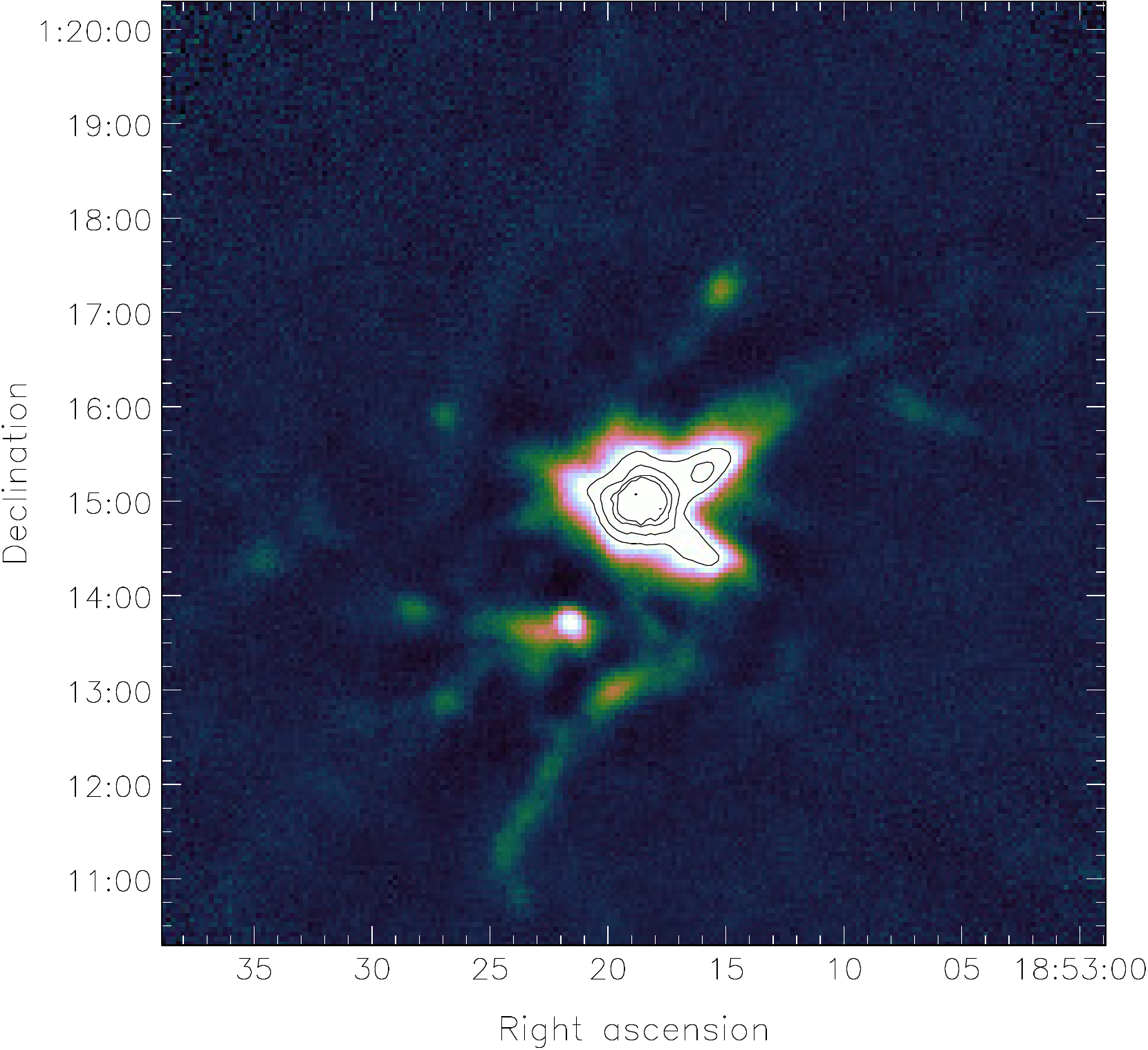}
\caption{SCUBA-2 observation of G34.3.}
\label{fig:g34}
\end{figure}

JCMT invested significant effort in automated data reduction based on
the ORAC-DR pipeline framework
\citep[][\ascl{1310.001}]{1999ASPC..172...11E,1999ASPC..172..171J,2005ASPC..347..585G,2008ASPC..394..565J,2015A&C.....9...40J}. In
many cases these automatically generated products were publication
quality, and thanks to a constantly updated calibration model, better
than what an inexperienced astronomer could be expected to achieve on
their own. Moreover with the advent of large bolometer arrays such as
SCUBA-2 \citep{2013MNRAS.430.2513H}, this data could be processed in
maps that resulted in image data that could be readily understood by
non-submm specialists, an example of which can be seen in \figref{fig:g34}.

The JCMT had in-house experience with setting up a data archive in the
``filing cabinet'' sense of allowing users to search and retrieve raw
data, but apart from a prototype involving the on-demand generation
of SCUBA data products \citep{2002ASPC..281..243J}, had not tackled
the integration of data processing with data
product distribution in a full science archive environment.
Indeed, distribution of publication-quality data became an issue of
the highest priority with the advent of the JCMT Legacy Survey
Programme \citep{2010HiA....15..797C,2008ASPC..394..450E} using the
SCUBA-2 and HARP/ACSIS \citep{2009MNRAS.399.1026B} instruments. Aside
from the normal desire to provide a uniformly reduced product to the
survey teams, the processing demands for this data required a
non-trivial IT infrastructure.
The complex iterative map-maker algorithm used to reduce SCUBA-2 data
\citep[SMURF;][\ascl{1310.007}]{2013MNRAS.430.2545C} was expected to
generate higher fidelity maps when more of an observation could be
fitted into memory at one time. It was estimated that at minimum a
machine with 64\,GB of RAM would be required (and 128\,GB is the
current recommendation) but circa 2008 machines of this size were not readily
available to the typical JCMT observer.
So there were intrinsic
reasons to have a JCMT Science Archive allowing the survey consortia
to download the processed products. Ultimately, usage of such a
standalone archive would be dominated by JCMT users retrieving their own data, or
after the proprietary period elapsed, other JCMT users working in the
same scientific areas who were explicitly searching for JCMT data.

JCMT formed a strong interest in going further, and exposing its
high-value data product to data-mining astronomers who would not have
a priori knowledge either of JCMT in particular or sub-mm astronomy in
general. To that end, the Virtual Observatory (VO) data discovery and publication protocols
seemed like a natural choice for reaching the large parts of the
astronomical community that were oblivious to its existence. VO
publication would also have the advantage of exposing the JCMT data
sets to workhorse tools that VO-savvy astronomers already used, such
as TOPCAT \citep[][\ascl{1101.010}]{2005ASPC..347...29T} and Aladin
\citep[][\ascl{1112.019}]{2005ASPC..347..193O}.

However, despite being convinced of the desirability of
leveraging the VO tools and services for JCMT data, the observatory
had the usual constraints of time and effort. The small Scientific
Computing Group was busy with supporting the entire non-hardware-controlling
software suite at both JCMT and UKIRT \citep[see e.g.,][with both
telescopes operated by the same organization]{2002SPIE.4844..321E,2011tfa..confE..42J},
as well as developing data reduction for new instruments, helping with
their commissioning,
and supporting the JCMT Legacy Surveys. The ability to develop a
VO-aware data centre and support the demands of the hoped-for
increased usage base was just not there.

What JCMT had, however, was a pre-existing collaboration with CADC,
which hosted the older JCMT data archive \citep{1997ASPC..125..397T}
for the benefit of the Canadian astronomical community, Canada being
one of the three international partners funding the JCMT (the other
two being the United Kingdom and the Netherlands). CADC had early
involvement in VO protocols \citep{2002ASPC..281...36S,2015ACCADC}, was a
productive developer and enthusiastic supporter of VO standards, and
was known to ``eat its own dog
food''\footnote{See \url{http://en.wikipedia.org/wiki/Eating_your_own_dog_food} and  \citet{2014arXiv1407.6463E} for more information.}
by using many of these
interfaces and services internally.

\section{Motivation: Data Centre}

\begin{figure*}[t]
\begin{center}
\includegraphics[width=0.8\textwidth]{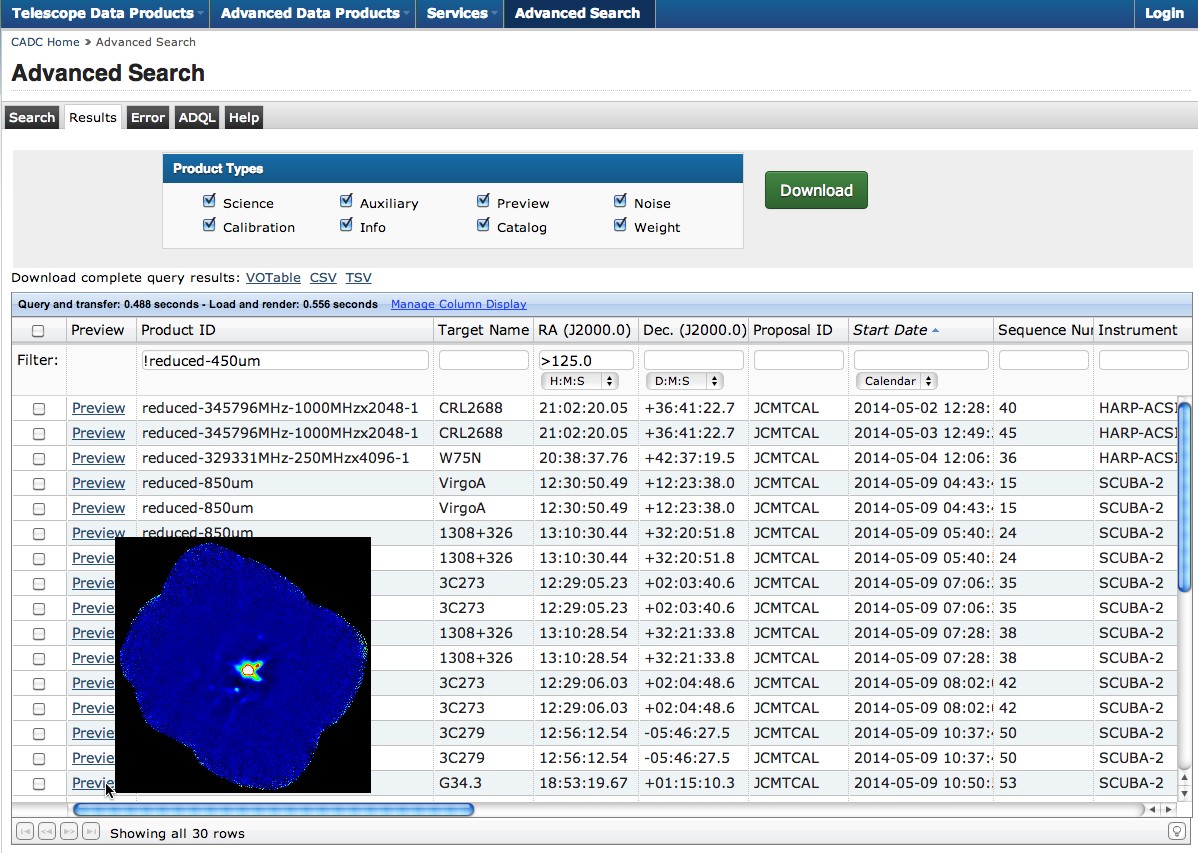}
\caption{AdvancedSearch Results.}
\label{fig:asresults}
\end{center}
\end{figure*}

CADC already had a varied collection of data from several
telescopes and space missions
\citep{1994ASPC...61..123C,2008SPIE.7016E..16G}. Keen to be able to
extend its holdings to new observatories and data sets while
requiring only a small and well-defined effort, CADC developed
the Common Archive Observation Model
\citep[CAOM:][]{2007ASPC..376..347D,2008ASPC..394..426D}.  CAOM defines an extensive and
versatile data model that classifies every data file using a common set of physical,
observational, organizational, and processing metadata.
This allows a generic VO search tool, such as AdvancedSearch, to search the
entire set of CADC archives for data relevant to a chosen target in the sky.

One of the main attractions of the JCMT data set was its significant
departure from many of the common forms of other astronomical data,
that predominantly came from optical and IR instrumentation.
Examples include:
\begin{itemize}
\item The ``photon energy'' axis for optical observations is normally described in wavelength
units like \AA{}ngstr\"{o}ms or microns, whereas most radio observations are defined in
frequency units like MHz and GHz.  To ingest and search for JCMT observations
it was necessary to enhance the tools to handle both wavelength and frequency units, with the
consequence that CADC interfaces now handle transparently most standard conversions
amongst frequency, energy and wavelength units.

\item At the start of the collaboration, most optical
data consisted of two dimensional RA/Dec images and sets of spectra.  Even at that time, JCMT data
came in RA/Dec, Galactic and offset co-ordinates, with up to 4 dimensions (2
spatial, wavelength and polarization).  The JCMT standard pipeline generates a
diverse set of products, including spectra, data cubes, maps, previews showing both spectral
and spatial images,  and catalogues for point sources, emission peaks and
clumps (extended regions of non-uniform emission).

\item Since most detector technologies only allow a photon to be
detected once, it can be safely assumed for optical instruments with multiple detectors
that the data products from different detectors will not overlap in WCS space.  The ability
at radio wavelengths to amplify the detected signal and feed it into multiple
spectrometers allows the output of the JCMT multi-subsystem spectrometer ACSIS
to include spectra and data cubes that overlap in a variety of ways, sometimes with
different frequency resolution, sometimes overlapping just at the ends of the spectra
to allow a much wider frequency coverage for a given frequency resolution than could be managed
by any single subsystem.
\end{itemize}

The JCMT therefore
provided an excellent stretch to the model, and continues to do so; if JCMT data could be
described in CAOM, CADC would be in the unprecedented position of
being able to accept almost any data set from future observatories
with minimal changes to their system.

Another advantage in working with JCMT on its datasets, was the high
level of completion and accuracy that JCMT provided in its
metadata. Even modern instruments on some older telescopes follow
metadata conventions established by the observatory long before the FITS
World Coordinate System (WCS) conventions were agreed upon.  At the
start of the collaboration, the CADC would assign an ``archive scientist'' to
each archive, whose job description included learning all the idiosyncrasies
of the observatory.   A major part of that effort
involved working around poor or incomplete metadata that made
astronomical data archiving problematic, especially if the observatory tended to
change their data products and headers without warning.  Maintaining a
proper ``Science Archive'' requires that both power users and
astronomers unfamiliar with an observatory's
internal conventions must be able to find and download science-ready data products
without mastering an arcane interface or guessing how to interpret the
metadata that it presents.
JCMT's dedication to high-fidelity
metadata and quick response in the rare case of problems made this an
attractive test data set.

The success of
this approach can be seen from the screen shot in \figref{fig:asresults},
which shows the reduced (Calibration Level 2) data from May 2014, sorted by
observation date, filtered to remove reduced-450\,$\upmu$m data (since the atmosphere
at 450\,$\upmu$m is often very opaque) and to include observations with RA $>$ 125.0 deg.
A pop-up preview of G34.3 is shown; clicking would bring up a larger version
of the preview in a new tab.  The \texttt{productID} column shows the kind of data that
can be downloaded for each selection, giving the product type (reduced data files in
this example) and basic wavelength information (filter for continuum observations,
rest frequency and spectrometer configuration for heterodyne observations).

\section{VO Standards Used in the JSA}

\begin{description}
\item[CAOM] : Common Archive Observation Model -- This is the data model
used in all archives at the CADC. It was designed to be a superset
of VO data models so that VO data models and services could be easily
implemented on top of CAOM. While CAOM is not a VO data model per se,
it was designed and is used as the metadata interface between archives
and standard VO data models. \citep{2007ASPC..376..347D,2013ASPC..475..159R}
\item[ObsCore] : Observation Data Model Core Components -- This VO data model
is designed to support data discovery specifically by supporting the
exact same queries to TAP services run by all data centres. In the JSA,
this is simply a view of CAOM as it contains a subset of CAOM metadata \citep{obscore}.
\item[SIA] : Simple Image Access -- Version 1.0 is an early VO service
interface that supports positional searching and retrieval of 2D
images \citep{siap}.
Version 2.0 \citep{siav2} is a new VO service interface that supports data discovery
of multi-dimensional datasets (images and data cubes) using the ObsCore
data model. Both of these are implemented using CAOM and TAP (below).
\item[TAP] : Table Access Protocol -- This VO service interface supports ad-hoc
querying of the CAOM metadata and standard views like ObsCore. All JSA
science data is discoverable through this interface \citep{tap,2014NandrekarHeinis201437}.
\item[ADQL] : Astronomical Data Query Language -- Queries to the TAP service
are formatted in ADQL, which is designed to closely resemble the popular SQL
syntax used by many relational database systems \citep{adql}.
\item[DataLink] : DataLink Service -- This VO service interface allows users
and client software to drill-down from discovered datasets to the list
of files to download and to services that can operate on the data. The
SIA-2.0 and TAP services use this interface to provide access to JSA
data files and services \citep{datalink}.
\item[AccessData] : Access Data Prototype -- This prototype VO service interface
allows users to perform cutouts on data files in a standard set of
world coordinates.
\item[CDP] : Credential Delegation Protocol -- This VO service interface enables
CADC services to call other services on behalf of the user so that the
correct identity and access rights are enforced. In the JSA, this allows
the user interface (AdvancedSearch) to pass the authenticated user
identity to the TAP service so that query results will include metadata
and access information for proprietary observations the user can access \citep{cdp}.
\item[VOTable] : Virtual Observatory Table Format -- This is a common tabular
format used to exchange metadata between clients and services. It is
the standard output format in SIA, TAP, and DataLink \citep{votable}.
\end{description}

\section{Evolution of the Data Flow}

\begin{figure*}[!ht]
\begin{center}
\includegraphics[width=0.75\textwidth]{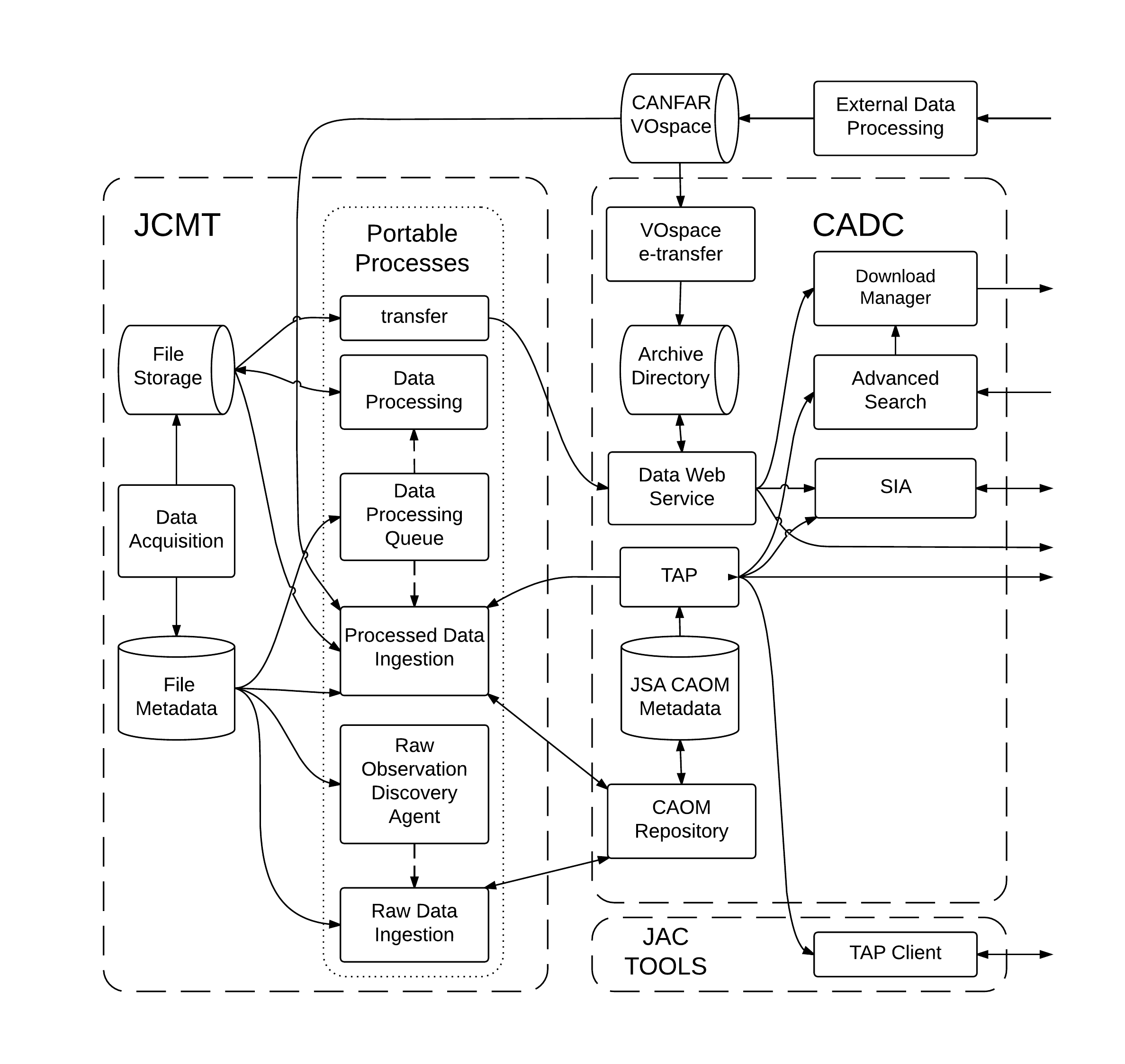}
\end{center}
\caption{Data and metadata flow through the JSA as it is intended to be in early 2015.  The JCMT and CADC processes are arranged in four columns, with the JCMT-specific processes on the left, ``portable processes'' (data processing, file transfer and metadata ingestion) in the dotted box in the centre-left, CADC processes in the centre-right, and client processes running on archive users machines on the right.  The ``vertical drums'' in the figure represent relational databases.  The ``horizontal drums'' represent file storage, but do not specify the technology used to implement the storage (disk drives at the JCMT, databases at the CADC and for the CANFAR VOSpace).  Where the software is developed, maintained and run by the JCMT/JAC or CADC, this is indicated by dashed container boxes.     Manual operations by JAC and CADC staff have been elided; user interactions are shown as arrows on the right side of the figure.}
\label{fig:jsadataflow}
\end{figure*}

The system that moves data from the JCMT to the CADC and on to our users has been under continuous development since the start of the collaboration.  \figref{fig:jsadataflow} shows the current development goal, which should have been attained by the time this paper is published.  Data files sent to be stored in the ``Archive Directory'' (AD) system at the CADC enter through the ``Data Web Service'' interface.  File metadata in the databases comprising the ``JSA CAOM Metadata'' system are managed using the ``CAOM Repository'' interface and can be read through the ``TAP'' service.  Similarly, users access data and metadata through the ``Data Web Service'' and TAP interfaces.  The use of a small number of well tested interfaces improves the reliability of the service and makes it easier to maintain on a limited budget.   Using the same interfaces that our users rely on ensures that problems are discovered and addressed quickly.

The system was initially quite different.  Before the advent of CAOM, every archive maintained a custom database.  Each file was stored in AD and ingested into the database as it arrived through e-transfer\footnote{For an introduction to the e-transfer system see \citet{2005ASPC..347..647M}}.  The JCMT supplied by replication a set of observatory databases that contained file metadata for raw data, and published an interface control document (ICD) describing the file headers in reduced data products. The JCMT committed itself to follow strict FITS standards for file headers and WCS, and for raw data reproduced a set of columns in the ``File Metadata'' database that was nearly identical  to the set of headers in the reduced data for single observations.   The CADC archive scientist was responsible for the design of software that read the metadata from the replicated database or from the reduced data headers.  Writing and maintaining the software to ingest the metadata into the ``JSA CAOM Metadata'' database required a team of software developers at the CADC.  The successful operation of this system required close collaboration of the JCMT with the JSA team at the CADC, with weekly progress videocons and regular (often annual) face-to-face meetings to discuss larger issues.  Although the system worked, it was cumbersome and expensive.  A leaner and more versatile system was clearly desirable.

The container labelled ``Portable Processes'' in \figref{fig:jsadataflow} illustrates how the leaner system was implemented.  The custom software for each archive was refactored into a set of simpler processes.  Data processing ran at the CADC for easy access to the stored data, but was developed and maintained by the JAC.  This encouraged a clean separation between the ``Data Processing Queue'' and  ``Data Processing'' itself.    The JSA was an early adopter of CAOM, which allowed raw and processed data ingestion to be factored out as separate processes.  Since raw data ingestion applies to whole observations, the ``Raw Data Discovery Agent'' verifies that all of the raw data for an observation is stored in AD before starting the ``Raw Data Ingestion'' process.  Originally, ``Processed Data Ingestion'' had its own discovery agent, but it is now controlled by the ``Data Processing Queue''.

The refactored system is quite modular and deployment is extremely
flexible. These processes were deployed at the CADC for most of the
last decade, but over the last year have migrated to the JAC.  Data
processing is currently run at the JAC using a queue system with
database tables similar to those used by CADC's original interface to
Sun Grid Engine. This has allowed the associated software to run with
minimal changes. The new system has a web interface which is tailored
to the JCMT, including a facility for in-house quality assurance. It
is anticipated that data processing might move onto a CANFAR Virtual
Machine in the near future and be orchestrated by the current queuing
system. Ingestion can now run on any node that can access the ``CAOM
Repository'', read existing metadata through the TAP service and
optionally access the ``File Metadata'' service at the JCMT.  This
extraordinary flexibility allows JCMT staff who best understand the
data to handle all data reduction and CADC staff who best understand
the archive to maintain those services.

\section{A Continuous Data Release Model}

Using CADC's data processing infrastructure and the capabilities of
JCMT/UKIRT's ORAC-DR automated data reduction, the JCMT Science
Archive adopted a model of continuous release
\citep{2011ASPC..442..203E}. As data was taken it was pushed for
reduction and was ingested at CADC in the same 24-hour period it was
observed. Thus, high-quality science products were published in the VO
as soon as the PI had access to them. Moreover, with every major
improvement in the data reduction software, data could be re-processed
and again immediately released.

Proprietary data goes into CAOM and becomes available via VO
interfaces almost immediately. Proprietary metadata and data
restrictions are enforced on all TAP queries and authentication will
permit authorized users to discover and download such data. Either
AdvancedSearch or direct TAP queries can be used by PIs and JCMT
legacy survey teams to find and download new data using this
authenticated access. For example, the Cosmology Legacy Survey team
\citep{2013MNRAS.432...53G} runs a script using the TAP interface to
keep track of new observations as they arrive in the archive. For year
2014, approximately 40~percent of all queries to the JSA came through
the TAP interface.

Continuous release made the VO publication mechanisms even more useful
than they are in the normal data discovery process, as product
availability is, from the point of view of the astronomer,
unpredictable rather than coming in fixed, scheduled, announced ``data
releases''. An interested user can therefore run regular automated TAP
queries with the expectation that newly-reduced data can appear from
their field of interest at any time.

\section{Post-Observatory}

Meanwhile, CADC was working on the Canadian Advanced Network for
Astronomical Research \citep[CANFAR;][]{2010SPIE.7740E..51G,2015ACCADC} project aiming to
support a cloud-like model for astronomical data reduction. The system
is based on giving the user a Virtual Machine (VM) that is then customized to
provide the appropriate software, environment and data access. The
user then defines a number of jobs that are serviced on a Condor
compute platform composed of customized VM copies.

This service has been of great utility to the Canadian astronomical
community dealing with large data volumes, with downloads of raw data from
the JSA to CANFAR processing nodes accounting for more than 40~percent of all
JSA raw data downloads in 2014.
The Gould Belt Legacy Survey \citep[GBS;][]{2007PASP..119..855W} make
use of use of CANFAR, and the GBS data
processing lifecycle is supported at every step by VO-compliant services.
Raw data is retrieved from VO-compliant discovery and delivery services,
that data is processed on the customized VMs provisioned on CANFAR, and
the resulting products are shared among survey members in VO-compliant
storage services using VOSpace \citep{vospace}. The total VOSpace
usage by the survey teams is currently approaching 1\,TB and this has
proven to be a critical part of the collaboration infrastructure when
dealing with teams spread over Canada, Hawaii and Europe.

The existence of the VOSpace system at CADC has also led to them
taking on the role of data publisher for JCMT science papers.
JSA data products and externally reduced products can be copied to a
VOSpace directory and associated with a
Digital Object Identifier. The first two data sets making use of this
functionality were \citet{2012MNRAS.424.3050W} and
\citet{2013ApJS..209....8D}.

\section{Extending VO for Radio Astronomy}

In the early days of the Virtual Observatory, the focus was
specifically on simple protocols \citep{siap,cone} to replace pre-existing
web services for image retrieval and cone search; with retrieval of
individual spectra coming somewhat later in VO developments
\citep{ssap,splatvo}. These were the pressing issues of the optical
community and this discussion dominated early protocol development.

Data cubes were seen as a task for the future as it was
felt that they were products that were not yet in the mainstream and
optical/IR instruments generating such cubes \citep[such as
the UIST IFU or TAURUS imaging Fabry-Perot spectrometer;][]{2004SPIE.5492.1160R,1982MNRAS.201..661A}
were seen as something of niche interest to be tackled later.
This was frustrating given that JCMT heterodyne
observations regularly generated cubes and with the arrival of ACSIS
in 2006, gigabyte data cubes were commonplace. There was no standard
available for making all these cubes available to the VO and it is
only recently \citep[e.g.,][]{2014AAS...22325505T} that a cube access
protocol has been approached with any seriousness, driven mainly, in
the USA, by ALMA and JWST developments \citep[e.g.,
MIRI;][]{2010SPIE.7731E..10W}. The proposed
recommendation for SIA-2.0 \citep{siav2}
will be able to handle the many data cubes generated by the JCMT
over the last two decades.

In Table~\ref{tab:cadcvo}, the line labeled ``TAP querying for
Spectra'' and ``TAP querying for Cubes''
indicate the number of 1-D spectra and data cubes in the JCMT collection.  These can
easily be found using the CADC AdvancedSearch interface, or directly using
a TAP query.  The full positional and photon energy WCS are provided
for these, even when the positional axes are degenerate.   SIA-2.0
should be able to find all of these data, once it has been implemented.

Another peculiarity of submillimetre data is the lack of point
sources. Most Galactic objects are extended and dust and gas from
large clouds, outflows and filamentary structures are missed by
standard source extraction algorithms such as SExtractor
\citep[][\ascl{1010.064}]{1996A&AS..117..393B}. Instead, algorithms
such as FellWalker \citep[][\ascl{1311.007}]{2015FW,2007ASPC..376..425B} and
Clumpfind \citep[][\ascl{1107.014}]{1994ApJ...428..693W}, which detect
source emission in irregularly shaped clumps, were used when
doing source finding. VO ConeSearch was not
set up for this eventuality and the best we could hope for was to
provide a catalogue that indicated the peak of the emission. To work
around this problem clump catalogues are generated with the clump
outline approximated by a polygon specified in STC-S format
\citep{2010ASPC..434..213B}. These outlines can then be retrieved
using TAP for analysis or plotting. This is certainly less convenient
for the end user than a clump equivalent of ConeSearch so we are
extending the facilities in GAIA
\citep[][\ascl{1403.024}]{2009ASPC..411..575D} to hide the TAP
interface. We hope a variant of ConeSearch will be developed that
works for extended irregular sources.
It should be sufficient for an enhanced ConeSearch query to return
the results as catalogues with STC-S columns representing the shape of
the object that matches, and for a match to be defined as an overlap
between the region specified by the caller and the region defining the
object. In this manner all existing ConeSearch services could simply
return objects with circular regions with size corresponding to the
point spread function.

\section{Current Status}

\begin{figure}[t]
\includegraphics[width=\columnwidth]{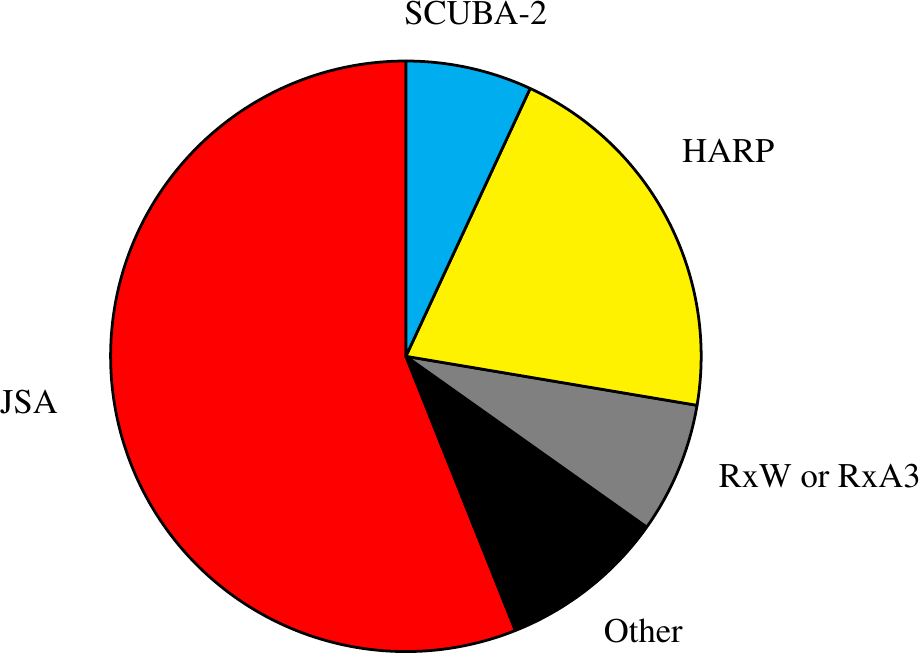}
\caption{Breakdown of the 405 JCMT refereed publications between 2010 and
    2013 indicating the fraction using data from the JCMT Science
    Archive. The remaining segments are from papers only using instrument
    data directly. Figure derived from \citet{2014SPIE9152-93}.}
\label{fig:jsapubs}
\end{figure}

\begin{table}
\caption{Data holdings in the JCMT Science Archive available over VO
  protocols as of 2014 November.}
\label{tab:cadcvo}
\begin{center}
\begin{tabular}{|l|r|}
\hline
Data model & Data sets available\\ \hline
TAP with CAOM (AdvancedSearch) & 1\,279\,617 \\
TAP with ObsCore & 1\,103\,787\\
TAP querying for cubes & 102\,392\\
TAP querying for spectra & 227\,839\\
SIA  & 335\,185\\ \hline
\end{tabular}
\end{center}
\end{table}

\begin{table}
\caption{Downloads of raw and reduced data from the JSA for the first
  11 months of 2014. 40~percent of the raw downloads are to CANFAR
  processing nodes. When interpreting the relative count of raw and
  processed files, note that SCUBA-2 generates 480 discrete data files
  every half hour, which may result in only two output maps (one for
  each wavelength, depending on tiling scheme).}
\label{tab:down}
\begin{center}
\begin{tabular}{|l|r|r|}
\hline
 & Number of files & Data volume / GB \\ \hline
 Processed  &     72\,730   &    1\,764 \\
 Raw           & 4\,427\,478    &  63\,611 \\  \hline
\end{tabular}
\end{center}
\end{table}

\figref{fig:jsapubs} demonstrates that between 2010 and 2013 more
than half of the refereed papers published containing JCMT data,
obtained data from the JSA. Table \ref{tab:cadcvo} provides the
current size of the data holdings accessible via a variety of VO
protocols, and table \ref{tab:down} provides details of how the
downloads from the JSA are split between raw and reduced data.

The collaboration has proven so successful that the opportunity was
taken to transfer the UKIRT raw data from the Cassegrain instruments
to CADC \citep{adassxxiii_P01}. It has been possible to re-use
the JSA processing infrastructure for UKIRT data processing as the
pipeline environment is identical \citep{2015A&C.....9...40J}.  Similarly, the ingestion
software initially developed for the JSA was easily adapted to ingest
data from several other CADC collections, including BLAST
(Balloon-borne Large Aperture Submillimeter Telescope, CGPS
(Canadian Galactic Plane Survey), IRIS (Improved Reprocessing
of the IRAS Survey), and VGPS (VLA Galactic Plane Survey).

The JSA data processing continues to be improved
\citep{2014JCMTN35..19J} and the current plan is to reduce all the
public HARP/ACSIS and SCUBA-2 data using an ``all-sky'' HEALPix
projection \citep{2005ApJ...622..759G,2014SPIE9152-93,2014JCMTN35..20B}. This processing
will also result in catalogue products that are specifically designed
to answer the question of whether the JCMT saw any emission in a
particular part of the sky. This is achieved by doing a two-pass
approach to clump
finding where first the emission outline is determined, and ultimately
represented by an STC-S polygon, and then the individual peaks are located
\citep{2014JCMTN35..21G}.

There is also an intent to expand the holdings of the JSA to include
heterodyne data taken in an older format by the DAS
\citep{1986SPIE..598..134B} and AOS-C backends. Data from those
instruments is being converted from the GSD format \citep{GSD1999} to the newer ACSIS format and this
allows all the standard processing infrastructure to be used to create
reduced data products and make them available to the VO for the first
time.

The JSA pioneered the use of CAOM at the CADC being implemented in both
CAOM-0.9 and CAOM-1.  The latest version, CAOM-2
\citep{2013ASPC..475..159R,2012ASPC..461..339D},  was released for general
use on 2014 May 1 and includes clarifications and improvements due to
lessons learned from the earlier models.
The metadata that is available for searching is richer, more complete, and easier
to understand than anything that has been available previously.
A full description of CAOM is in the early stages of preparation, but the earlier
references cited above still describe the core philosophy of the design, and the
current database schema can be found online\footnote{\url{http://www.cadc-ccda.hia-iha.nrc-cnrc.gc.ca/caom2/}}.

\section{Lessons Learned}

The JCMT Science Archive collaboration was a high successful foray
into VO publication via an observatory-data centre collaboration.

Elements that we believe led to this success:

\begin{itemize}
\item VO publication was a common goal with significant organizational
  buy-in for both parties from the start, and was a primary technical
  goal of the collaboration rather than an afterthought.

\item Within that shared vision, there was a clear division of
  expertise and responsibilities for each side, allowing each
  organization to focus on its proximate technical goals. Both
  organizations had ``skin in the game'' that was served by the
  technical work undertaken, which allowed this work to be carried out
  without any kind of external agency funding (each institution
  supported its own share of the work out of its normal budgetary
  process).

\item Each organization worked from a position of strength based on an
  advanced, robust and mature software architecture, allowing
  development to focus on new functionality and interfaces between the
  two systems. This minimized the communication overheads commonly
  associated with distributed projects.

\item The role of ``data engineer'' responsible for developing software to
ingest new data into a CAOM archive no longer requires special privileges
at the CADC.  It does require an expert
knowledgable about both the CAOM model and the products generated
by the data reduction system, but the tools developed for CAOM
allow this role to be assigned to the best available expert regardless of their
location or institutional association.  Thus, for UKIRT, the Joint Astronomy Centre has been able to
assign one of their own staff to this role, and for the JSA a retired CADC staff
member currently fills the role.

\item There was a high level of pre-existing trust between the two
  groups from their previous relationship leading to minimal need for
  contractual language or management oversight. Indeed the entire
  collaboration's only official governance document was a two paragraph
  memorandum of understanding.
\end{itemize}

\section{Recommendations}

In the general case, for observatories that do not understand the
mechanisms or benefits of VO publication, collaboration with a
motivated VO-involved data centre that has the appropriate
infrastructure and keeps up to date with the IVOA standards process is
a far more effective choice than trying to develop those capabilities
in-house, especially since there seems to be confusion in the
observatory community as to what ``VO publication'' involves and what
are the merits of doing it.

However, in order to be able to properly leverage the capability of a
modern multi-mission data centre, a fanatical devotion to correct and
complete metadata should be considered a pre-requisite.

Good communications within the team of collaborators is essential.  Regular
weekly or bi-weekly teleconferences and occasional face-to-face meetings
have been important to keeping everyone aware of issues and working to
common purposes.

\section*{Acknowledgments}

The James Clerk Maxwell Telescope has historically been operated by
the Joint Astronomy Centre on behalf of the Science and Technology
Facilities Council of the United Kingdom, the National Research
Council of Canada and the Netherlands Organisation for Scientific
Research.  Additional funds for the construction of SCUBA-2 were
provided by the Canada Foundation for Innovation.
The Canadian Astronomy Data Centre is operated by the
National Research Council of Canada with the support of the Canadian
Space Agency. This research has made use of NASA's Astrophysics
Data System.

\end{document}